\documentclass{article}

\PassOptionsToPackage{numbers, compress}{natbib}

\usepackage[preprint]{neurips_2022}

\usepackage[utf8]{inputenc} 
\usepackage[T1]{fontenc}    
\usepackage{hyperref}       
\usepackage{url}            
\usepackage{booktabs}       
\usepackage{amsfonts}       
\usepackage{nicefrac}       
\usepackage{microtype}      
\usepackage{xcolor}         

\usepackage{bm}
\usepackage{amsmath}
\usepackage[utf8]{inputenc}
\usepackage{graphicx}
\usepackage{algpseudocode}
\usepackage{algorithm}
\usepackage{amsmath, amsfonts}
\newcommand{\T}[0]{\beta}
\usepackage{amsthm}

\newtheorem{theorem}{Theorem}

\title{Continuously-Tempered PDMP samplers}
\author{%
   Matthew Sutton \\
    Centre for Data Science\\
  Queensland University of Technology\\
  \texttt{matt.sutton@qut.edu.au} \\
   \And
  Robert Salomone \\
  Centre for Data Science\\
  Queensland University of Technology\\
  \texttt{robert.salomone@qut.edu.au} \\
  \And
  Augustin Chevallier \thanks{The authors acknowledge funding through EPSRC grants 	EP/R018561/1 and EP/R034710/1} \\
  Department of Mathematics and Statistics\\
  Lancaster University\\
  \texttt{a.chevallier@lancaster.ac.uk} \\
  \And
   Paul Fearnhead \\
    Department of Mathematics and Statistics\\
  Lancaster University\\
  \texttt{p.fearnhead@lancaster.ac.uk} \\
}
\begin{document}

\maketitle

\begin{abstract}
  New sampling algorithms  based on simulating continuous-time stochastic processes called piece-wise deterministic Markov processes (PDMPs) have shown considerable promise. However, these methods can struggle to sample from multi-modal or heavy-tailed distributions. We show how tempering ideas can improve the mixing of PDMPs in such cases. We introduce an extended distribution defined over the state of the posterior distribution and an inverse temperature, which interpolates between a tractable distribution when the inverse temperature is 0 and the posterior when the inverse temperature is 1. The marginal distribution of the inverse temperature is a mixture of a continuous distribution on $[0,1)$ and a point mass at 1: which means that we obtain samples when the inverse temperature is 1, and these are draws from the posterior, but sampling algorithms will also explore distributions at lower temperatures which will improve mixing. We show how PDMPs, and particularly the Zig-Zag sampler, can be implemented to sample from such an extended distribution. The resulting algorithm is easy to implement and we show empirically that it can outperform existing PDMP-based samplers on challenging multimodal posteriors.
\end{abstract}

\section{Introduction}
Recently there has been considerable interest in developing sampling algorithms based on simulating continuous time stochastic processes called a piecewise deterministic processes (PDMPs) \cite{Davis1993}. These sampling algorithms are qualitatively similar to Hamiltonian Monte Carlo algorithms \cite{neal2011mcmc}. To simulate from some targst distribution $\pi(\bm{x})$, they work with an augmented state-space $(\bm{x},\bm{v})$, where the $\bm{x}$ component can be viewed as position, and the $\bm{v}$ component as a velocity. The motivation for this is that these processes will encourage exploration of $\pi(\bm{x})$ by simulating continuous velocity paths in between random event times at which the velocity changes. Examples of such sampling algorithms include the Bouncy Particle Sampler \cite{peters2012rejection,Bouchard-Cote2018}, the Zig-Zag algorithm \cite{Bierkens2019}, the Coordinate Sampler \cite{wu2020coordinate} and the Boomerang Sampler \cite{bierkens2020boomerang}. See \cite{fearnhead2018piecewise} for an overview of these methods. Importantly, it is simple to write down the event rate, and how the velocity should change at each event, to ensure these PDMPs have $\pi$ as their invariant distribution. These depend on $\pi$ only through the gradient of $\log \pi$, and thus $\pi$ only needs to be known up to a constant of proportionality.

These PDMP samplers have a number of advantages, including non-reversible dynamics (which is known to improve mixing relative to reversible processes \cite{diaconis2000analysis,bierkens2016non}), and the ability to reduce computation-per-iteration by either leveraging sparsity structure in the model \cite{Bouchard-Cote2018,sutton2021concave} or using only sub-samples of the data to approximate the log-likelihood at each iteration (whilst still guaranteeing sampling from the target \cite{Bierkens2019}). However, like other MCMC algorithms, particularly those that use gradient information, these PDMP samplers can struggle to mix for multi-modal target distributions, or for heavy-tailed targets \cite{Vasdekis2021}.

One of the more successful techniques for enabling an MCMC algorithm to sample from challenging, e.g. multi-modal, target distributions is to use tempering. There are various forms of tempering, but each is based on defining either a discrete set or continuum of distributions that interpolate between a distribution that is simple to sample from (viewed as at high temperature) and the target distribution (at a low temperature). The idea is that allowing mixing across this set will improve mixing, as moving between modes will be easier for the distributions at higher temperatures. Examples of such algorithms include parallel tempering \cite{swendsen1986replica}, simulated tempering \cite{marinari1992simulated}, and continuous tempering \cite{Graham2017}.

In this paper we show how continuous tempering ideas can be used with PDMP samplers. We have chosen {\em continuous} tempering, as opposed to the alternative tempering approaches, as it is the method that can most naturally benefit from the continuous-time nature of PDMP dynamics. To the best of our knowledge, this is the first attempt at using tempering ideas to improve PDMP samplers. The general idea is to define a joint distribution on the state of the PDMP, $\bm{z}=(\bm{x},\bm{v})$, and the inverse temperature, $\beta$. The target distribution of interest is the $\bm{x}$-marginal of this joint distribution when $\beta=1$. We define the joint distribution so that it has a point mass at $\beta=1$ -- thus simulating from it will lead to a proportion of the resulting samples being from the target. We then use a PDMP sampler to simulate from this joint distribution. Constructing the appropriate dynamics of the PDMP sampler is non-trivial in this case as we have to deal with its behaviour as it transitions from the continuous distribution on $\T\in[0,1)$ and the point mass at $\beta=1$. We use recent ideas for PDMP samplers with discontinuities \cite{chevallier2021pdmp,chevallier2020reversible} to solve this challenge.

While we express ideas based on and similar to \cite{Graham2017}, a key difference is the inclusion of a point-mass at $\T=1$, this means we can obtain samples from $\pi(\bm{x})$ rather than having to resort to importance sampling to correct samples drawn at different temperatures. It is easy to introduce a point mass into the dynamics of the PDMP sampler due to its continuous sample paths: we simulate paths for $\T\in[0,1)$ until the process hits $\T=1$ -- we then simulate paths with $\T$ fixed to 1 for an exponential period of time before returning to $\T\in[0,1)$. By comparison, using a point-mass within a Hamiltonian Monte Carlo sampler is difficult as the discretised sample paths will not hit $\T=1$ with probability 1. \cite{Yao2020} consider an {\em indirect} approach to sampling with a pointmass at $\T=1$ using a continuous link function. Our approach is distinct and more direct, allowing one to exploit the unique advantages of PDMP samplers --- such as the ability to perform subsampling without altering the ergodic distribution and application to sampling transdimensional distributions.

The benefits of introducing the point mass at $\beta=1$  are numerically investigated, and practical considerations related to choosing tuning parameters to encourage a desired proportion of time at the target distribution are presented (Section \ref{sec:choose_kappa}). We find that, analogously to standard methods in discrete-time, the tempered counterparts of PDMP samplers outperform vanilla PDMP on challenging sampling problems.

\section{The Zig-Zag sampler}

For the purposes of brevity and ease of exposition, we focus specifically on a continuous-tempered version of the Zig-Zag sampler. However, we stress that the underlying ideas can easily be applied to {\em any} PDMP sampler, for example, the Bouncy Particle Sampler \cite{Bouchard-Cote2018} or the Boomerang Sampler \cite{bierkens2020boomerang}; see the comments at the end of Section \ref{sec:CTZZ}.

Consider the problem of sampling from a target density defined for $\bm{x}\in \mathcal{X} := \mathbb{R}^d$ by
$$
\pi(\bm{x}) = \frac{1}{Z}\exp(-U(\bm{x}))
$$
where $U : \mathcal{X}\rightarrow \mathbb{R}$ is a continuous differentiable function referred to as the potential and $Z$ is the, potentially unknown, normalising constant $Z = \int_{\mathcal{X}}\exp(-U(\bm{x})) d\bm{x}$. We will denote the un-normalised target density by $q$, so $\pi(\bm{x})=q(\bm{x})/Z$.

The {\em Zig-Zag} process is a continuous-time piece-wise deterministic Markov process, which can be defined so as to have $\pi(\bm{x})$ as its invariant distribution. The process is defined on an extended state-space that can be viewed as consisting of a position, $\bm{x}$ and a velocity component, $\bm{v}$. For the Zig-Zag process, the velocity is restricted to be $\pm 1$ in each axis direction. Thus the extended space is  $E = \mathbb{R}^d \times \{-1,1\}^d$.  We write $\bm{z} = (\bm{x},\bm{v})$ for $\bm{z} \in E$ from here on. We will use subscripts to denote time, and superscripts to denote components. So $\bm{z}_t$ will be the state at time $t$, while $\bm{x}^i_t$ will be the $i$th component of the position at time $t$.

For an event at time $t$ we use the notation $\bm{z}_{t-}$ to be the state immediately before the event, and $\bm{z}_t$ the state immediately after it.
The dynamics of the Zig-Zag process are deterministic between a set of random event times. At each event time the direction of one component of the velocity is switched. The deterministic dynamics are specified by a constant velocity model. 
So, if there are no events between times $t$ and $t+h$, for $h>0$, the change of state is given by $\bm{z}_{t+h}=(\bm{x}_{t+h},\bm{v}_{t+h})=(\bm{x}_t+h\bm{v}_t,\bm{v}_t).$

The events occur with a rate that depends on the current state. For the Zig-Zag process we have $d$ types of event, each of which results in the flipping of one of the $d$ components of the velocity process. To ensure that we have $\pi(\bm{x})$ as the $\bm{x}$-marginal of the process's invariant distribution, these rates are defined to be, for $i=1,\ldots,d$,
\[
    \lambda_i(\bm{x}_t, \bm{v}_t) = \max(0,\bm{v}_t^i \partial_{\bm{x}^i} U(\bm{x})),
\]
with the transition at the corresponding event being that $\bm{v}_{t}^i=-\bm{v}_{t-}^i$, and all other elements of the state are unchanged.  

Pseudo-code for simulating the Zig-Zag process is given in Algorithm \ref{alg:ZZ}. When we simulate such a process, the output of the algorithm is the set of event times, positions and velocities after the event times. This set is known as the PDMP skeleton and with the deterministic dynamics of the process it defines the continuous time path of the process. We can use such a simulated path to give us draws from $\pi(\bm{x})$, by discarding some suitably chosen initial path time as burn-in, and then evaluating the $\bm{x}$ component of the process at a set of evenly spaced discrete time-points (see e.g. \cite{fearnhead2018piecewise} for alternative approaches that use the continuous-time paths).

\begin{algorithm}
\caption{Zig-Zag algorithm \label{alg:ZZ}}
\begin{algorithmic}[1]
\State {\bf{Inputs:}} initial state
$(\bm{x}_{t_0}, \bm{v}_{t_0})$ and number of simulated events $K$
\State $t_0 \gets 0$
\For{k $\in 1,\dots,K$ } 
    \State Simulate an event-time $\tau_i$ for each rate $\lambda_i(\bm{x}_t, \bm{v}_t)$
    \State $i^* \gets \text{argmin}_i\{\tau_i\}$ and $t_k \gets t_{k-1} + \tau_{i^*}$
    \State$\bm{x}_{t_k} \gets \bm{x}_{t_{k-1}} + \tau_{i^*}\bm{v}_{t_{k-1}}$ \Comment{\textit{Update position}}
    \State $\bm{v}_{t_k}^{i^*} \gets -\bm{v}_{t_{k-1}}^{i^*}$ \Comment{\textit{Update velocity}}
\EndFor
\State {\bf{Outputs:}} Zig-Zag skeleton $\{t_k, \bm{x}_{t_k}, \bm{v}_{t_k}\}$.
\end{algorithmic}
\end{algorithm}

\section{Continuously-tempering with a point mass}

\subsection{Continuous-tempering}

Continuous-tempering is an approach to improve mixing of MCMC and related sampling algorithms. It introduces a second distribution $\pi_0$, called the {\em base} distribution, which is assumed to have a density $\pi_0(\bm{x}) = \frac{q_0(\bm{x})}{Z_0}$. The idea is that this density will be simple to simulate from.  For any $\T\in[0,1]$ it is now possible to define a distribution which interpolates between $\pi$ and $\pi_0$ in that its density is $\pi(\bm{x})^\T\pi_0(\bm{x})^{(1-\T)}$. Continuous-tempering then defines a joint distribution on $\T$ and $\bm{x}$. This distribution has un-normalised density  $q(\bm{x},\T)$,  on $\mathbb{R}^d \times [0,1]$, defined to be 
\[
q(\bm{x},\T) = q_0(\bm{x})^{1-\T}q(\bm{x})^\T.
\]
In practice tempering methods often introduce a prior, $p(\T)$ on $\T\in[0,1]$, and sample from the distribution proportional to $p(\T)q(\bm{x},\T)$. The idea of introducing the prior is that it can be tuned to give a marginal distribution on $\T$ that has reasonable mass across all of the interval $[0,1]$. Without a prior, the distribution $q$ is likely to put almost all mass on $\T$ close to 0 or on $\T$ close to 1.

By construction, conditional on satisfying $\T=1$, samples drawn from $q(\bm{x}, \T)$ are distributed according to $\pi$. However, if we use a continuous prior, samples will almost-surely {\em not} lie in that set, and we have to use importance sampling to correct for $\T\neq 1$. This leads to a weighted sample from $\pi$, with the Monte Carlo accuracy of the resulting approach depending greatly on the variability of the weights introduced by importance sampling.

To overcome this issue, we use a prior $p(\T)$ that introduces a point mass at $\T=1$. 
Specifically, for $\alpha \in \mathbb{R}$, and an arbitrary probability density function $\kappa$ on the interval $[0,1)$, we define 
$$
p(\T) = (1-\alpha)\kappa(\T)\mathbf{1}_{(0\leq \T<1)} + \alpha\kappa(\T)\mathbf{1}_{(\T=1)} .
$$

With the above prior, we can define the augmented target
\begin{equation}    \label{eq:extendedmeasure}
\omega(\mbox{d}\bm{x}, \mbox{d}\T) \propto q(\bm{x},\T)(1-\alpha) \kappa(\T) \mbox{d}\bm{x} \mbox{d}\T +  q(\bm{x})\alpha\kappa(\T) \mbox{d}\bm{x} \delta_{\T=1}.
\end{equation} 

We can extend the Zig-Zag process to sample from such a distribution. The next subsection describes the extension. The usual advantages of subsampling and transdimensional sampling using PDMPs apply for the tempered extension. 

\subsection{Continuously-tempered Zig-Zag} \label{sec:CTZZ}

Unlike most MCMC samplers, Zig-Zag is a continuous-time process. This makes it simpler to deal with the point-mass at $\T=1$, using ideas from \cite{chevallier2020reversible}: we simulate paths for $\T\in[0,1)$ until $\T$ hits the boundary at $\T=1$; when this happens, we set the velocity component for $\T$, denoted as $\bm{v}_t^{d+1}$, to zero so that  the Zig-Zag process explores the distribution $\omega$ restricted to $\T=1$; the Zig-Zag process stays with $\T=1$ until the first event in a new Poisson process, at which point we set the velocity component for $\T$ to $-1$ to allow exploration of the distribution for $\T\in[0,1)$.  We define the rate of the events when we leave $\T=1$ to be  $\eta$. Algorithm \ref{alg:ZZ-CT} provides a sketch of the relevant modifications of the standard Zig-Zag process for events concerning the inverse temperature variable $\beta$. 

\begin{algorithm}
\caption{Tempered Zig-Zag algorithm \label{alg:ZZ-CT}}
\begin{algorithmic}[1]
\State Run Zig-Zag for $\T_t\in[0,1)$ until first time $t$ at which trajectory hits $\T=1$
    \State $\bm{v}^{d+1}_t \gets 0$ \Comment{\textit{Kill the velocity component for $\T$}}
    \State Simulate an event-time $\tau$ with rate $\eta$ 
    \State Run the Zig-Zag process with $\T=1$, for time $\tau$ \Comment{\textit{Sample $\pi$}}
    \State $\bm{v}^{d+1}_{t+\tau} \gets -1$ \Comment{\textit{Reintroduce $\T$}}
    \State Goto Step 1.
\end{algorithmic}
\end{algorithm}

An appropriate choice of rate to ensure that the resulting algorithm generates a process with $\omega$ as its limiting distribution is given in the following result (the proof of is located in the supplement).

\begin{theorem}
    Assuming $\kappa(\T)$ is continuous and the rate function in Algorithm \ref{alg:ZZ-CT} is chosen as
    \[
        \eta = \frac{1- \alpha}{2\alpha},
    \]
    Then, the resulting Zig-Zag process is $\omega$-ergodic. 
\end{theorem}

The theorem statement above considers the case where $\kappa$ is continuous but we note in the proof that more general choice is possible with a slight adjustment to the rate. 

Importantly, the rate $\eta(\bm{x})$ is constant, and thus simulating the time at which we transition from $\T=1$ to $\T<1$ is independent of the path of the process and simulating the event time is simple.
While we have described the use of continuous tempering as an extension of the Zig-Zag process, the same construction readily extends to other PDMP samplers --- one simply replaces the Zig-Zag dynamics for $\beta\in[0,1)$ and $\beta=1$ in Algorithm \ref{alg:ZZ-CT} with the dynamics of the corresponding PDMP sampler.

\subsection{Importance sampling estimator}
\label{sec:is_estimator}
By construction, continuously-tempered Zig-Zag will spend a substantial amount of time in states with $\T=1$, and thus the states at those times can give draws from $\pi(\bm{x})$. We can use importance sampling ideas from  \cite{Graham2017} to re-weight samples when $0\le \T<1$ to give weighted samples from $\pi(\bm{x})$; but this only applies if $\kappa(\T)\propto \xi^{1-\T}$ for some constant $\xi$.

The idea of importance sampling is that, for $\T\in[0,1)$ we can marginalise out $\T$ from $\omega$, and the marginal distribution for $\bm{x}$ is
\[
\omega_{[0,1)}(\bm{x}) = \int_0^1 \kappa(\T) q(\bm{x}) \left(\frac{q_0(\bm{x})}{q(\bm{x})}\right)^{1-\T} \mbox{d}\T=
q(\bm{x}) \int_0^1 \left(\frac{ \xi q_0(\bm{x})}{q(\bm{x})}\right)^{1-\T} \mbox{d}\T. 
\]
Defining $\Delta(\bm{x}) = \log q_0(\bm{x}) + \log \xi - \log q(\bm{x})$, the above is equal to
\[
q(\bm{x}) \int_0^1 \exp\{ (1-\T) \Delta(\bm{x}) \}  \mbox{d}\T=\exp\{\Delta(\bm{x})\}\Delta(\bm{x})^{-1} (1-\exp\{-\Delta(\bm{x})\})=\frac{\exp\{\Delta(\bm{x})\}-1}{\Delta(\bm{x})}.
\]
Thus if we have this as our proposal distribution, then the corresponding importance sampling weights will be $w(\bm{x})= \Delta(\bm{x})/[\exp\{\Delta(\bm{x})\}-1]$. This will involve additional post-sampling computation of the importance weights at the samples taken from the Zig-Zag trajectory. 

\subsection{Calibration of $\kappa(\T)$}\label{sec:choose_kappa} 

The efficiency of continuous-tempering depends on an appropriate choice of $\kappa$ and $\alpha$. Ideally, they should be chosen so to balance a non-negligible amount of time at $\T=1$, while simultaneously allowing occasional excursions to lower inverse temperatures to help mix, e.g. between modes.

For $\T \in [0,1]$ define,
$Z(\T) = \int_{\mathbb{R}^d} q(\bm{x},\T) d\bm{x}$.
The induced $\beta$-marginal of $\omega$ for $\T_0 \in [0,1)$ is 

$$
\omega(\T_0) = \frac{(1-\alpha)\kappa(\T_0) Z(\T_0)}{(1-\alpha) \int_0^1 \kappa(\T)Z(\T) d\T + \alpha \kappa(1)Z(1)}.
$$

The above suggests that an appropriate choice is $\kappa(\T)\propto Z(\T)^{-1}$, as this would both induce $\beta$ to be marginally {\em uniform} under $\omega$ for $\T\in[0,1)$, and cause the parameter $\alpha$ to directly represent the probability that $\T=1$ under $\omega$. When not using the important sampling strategy, the proposed approach is thus to choose $\kappa(\T)\propto\widehat{Z}(\T)^{-1}$, where $\widehat{Z}(\T)= \exp\left\{\sum_{k=0}^{m-1} a_k \T^k \right\}$ for some coefficients $\{a_k\}_{k=0}^{m-1}$ that are estimated by regressing point estimates of $\log Z(\beta)$ obtained via the path sampling identity. In practice,  an estimate $\widehat{Z}(\beta)$ can be obtained from a pilot run of the algorithm with $\alpha\ll 1$ or using methods described in \cite{Gelman1998}.

\section{Numerical experiments}
\subsection{Mixture of Gaussians}
In our first example, the target corresponds to a mixture of Gaussian distributions with equal weights and variances so our target has unnormalised density
$$
q(\bm{x}) = \sum_{i=1}^K\exp\left(-\frac{1}{2\sigma^2}(\bm{x} - \bm{\mu}_i)^\top(\bm{x} - \bm{\mu}_i)\right),
$$
where $K=5$, $\sigma^2 = 0.2$ and $\{\bm{\mu}_1,...,\bm{\mu}_5\}$ were generated uniformly on the region $[0,10]\times[0,10]$ and are given in the supplementary material. Figure \ref{fig:gmix_results} plots the PDMP trajectories for the tempered and untempered Zig-Zag sampler, in addition to the trajectories for inverse temperature fixed at $\T = 1$. For this example we choose $q_0(x)$ as a Gaussian $\mathcal{N}(\nu, \Sigma)$ centred at $\nu = (5,5)^T$ with covariance matrix $\Sigma = 2\mathbf{I}_2$. 

\begin{figure}[ht]
    \centering
    \includegraphics[width=.8\textwidth]{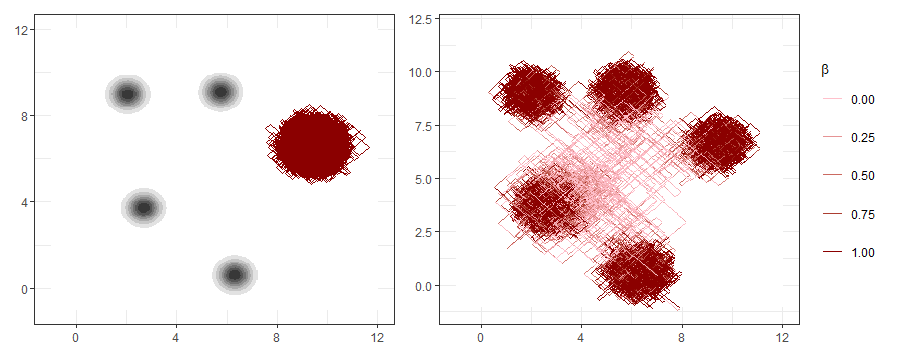}
    \caption{Trajectories of the Zig-Zag process simulated for 30,000 events for a multi-modal Gaussian mixture model, Zig-Zag (left) and continuously tempered Zig-Zag (right) with $\alpha = 0.7$.}
    \label{fig:gmix_results}
\end{figure}

\begin{table}[ht]
\caption{Recovery of first two moments of a Gaussian mixture (averaged over 20 replications).}
\label{tab:gmix_results}
\begin{tabular}{@{\extracolsep{5pt}} ccc|cccc|c} 
\\[-1.8ex]\hline 
\hline \\[-1.8ex] 
&&& \multicolumn{4}{c|}{Root-mean-square error (RMSE)} & Thinning \\
Method & $\alpha$ & $\omega(\T=1)$ & $\mathbb{E} [X_1] $ & $\mathbb{E}[X_2]$ & $\mathbb{E}[X_1^2]$ & $\mathbb{E}[X_2^2]$ & efficiency \\ 
\hline \\[-1.8ex] 
Zig-Zag & $1$ & $1$ & $2.557$ & $2.740$ & $30.577$ & $31.761$ & $0.057$ \\ 
Zig-Zag CT & $0.8$ & $0.789$ & $0.650$ & $0.741$ & $7.898$ & $7.182$ & $0.080$ \\ 
 & $0.7$ & $0.703$ & $0.399$ & $0.683$ & $4.563$ & $6.418$ & $0.090$ \\ 
 & $0.5$ & $0.499$ & $0.329$ & $0.539$ & $4.199$ & $4.930$ & $0.114$ \\ 
 & $0.3$ & $0.302$ & $0.304$ & $0.453$ & $\bm{3.216}$ & $\bm{4.155}$ & $0.139$ \\ 
 & $0.2$ & $0.197$ & $\bm{0.294}$ & $0.472$ & $3.756$ & $4.617$ & $0.153$ \\ 
 & $0.1$ & $0.097$ & $0.349$ & $\bm{0.389}$ & $3.987$ & $4.198$ & $0.167$ \\ 
Zig-Zag CT (IS) &   $0$ & $0$ & $0.483$ & $0.472$ & $5.567$ & $5.030$ & $\bm{0.301}$\\
\hline \\[-1.8ex] 
\end{tabular} 
\end{table}

Table \ref{tab:gmix_results} displays the root-mean-square-error (RMSE) of the Monte Carlo estimates from the Zig-Zag and tempered Zig-Zag averaged over 20 runs. All methods were simulated for 50,000 event-times with the first 40\% used as burnin in the standard Zig-Zag and used for both burn-in and estimating the polynomial $\kappa(\T)$ in the tempered samplers. The standard Zig-Zag is not able to explore the multiple modes yielding worse estimates of the first and second moments. We also compare to a direct Zig-Zag analogue of the continuously tempered HMC algorithm \cite{Graham2017} where no mass is given to $\T = 1$ and $\kappa(\T)\propto 1$, the estimator is based on importance sampling as defined in Section \ref{sec:is_estimator}. Further details and boxplots showing variability of the estimated moments are given in the supplementary material. 

We find that the tuning procedure yielded precise control over the time spent at $\T=1$, and that for $\alpha$ between 0.1 and 0.3 the results are similar. Tempering with a point mass was found to give more efficient estimates of the first and second moments. This may be because the importance sampling estimate spends more time at $\T\approx0$. In addition to exhibiting lower RMSEs, the tempered versions of the Zig-Zag sampler have better computational properties. The thinning efficiency, measured as the proportion of proposals that result in an event-time simulation is $\approx .06$ at $\T=1$ but improves significantly when the sampler can transition to lower temperatures. While the importance sampling estimator for $\T=0$ has the best thinning efficiency it requires additional post processing to evaluate the importance sampling weights.

\subsection{Transdimensional example}\label{sec_exptrans}
We apply the tempered Zig-Zag to the challenge of sampling a transdimensional distribution. Such target distributions arise naturally in variable selection problems. The resulting target (posterior) distribution is a discrete mixture of $2^p$ models, where $p$ is the number of variables in the dataset.

Such a setting produces a challenge, as typical samplers such as HMC do not extend naturally to such spaces. On the other hand, PDMPs have recently been extended to sample  transdimensional distributions \cite{chevallier2020reversible, chevallier2021pdmp, bierkens2021sticky}. This extension employs within-model gradient information to efficiently explore the space and jumps between models when a parameter hits zero. The approach is beneficial as an informative likelihood function's gradient will direct less informative variables towards zero and more informative ones away from zero, aiding exploration of the sampler. 

This example explores how tempering a target with a point mass can improve performance over the standard transdimensional Zig-Zag process. Specifically, we define a family of example problems of increasing difficulty in the sense that for higher values of an underlying parameter $m$, separation between the mode of the slab component and the spike at zero is increased. Allow
\begin{align}
\label{eq:spikeslab}
    q(\bm{x}) = \prod_{i=1}^2(w\phi(x_i; m, \sigma^2) + (1-w)\delta_0(x_i)),
\end{align}
where $\phi(x_i; m, \sigma^2)$ is the normal density and $w = 0.5$ is the probability of a variable being included. We fix $\sigma^2=0.5$ and increase $m$ from $m=0$ to $m=4$. To enable tempering, define
$$
q(\bm{x}, \T) = \prod_{i=1}^2(w\phi(x_i; m\T, \sigma^2) + (1-w)\delta_0(x_i)),
$$
where at $\T=0$ the spike and slab is centred at zero, encouraging variables to enter and exit the model. While this is a somewhat artificial example, it is important to note that it encompasses precisely the situations which transdimensional PDMPs will find challenging, and allows us to explore robustness for increasing levels of pathology in the purest setting possible.

\begin{figure}[ht]
    \centering
    \includegraphics[width=.9\textwidth]{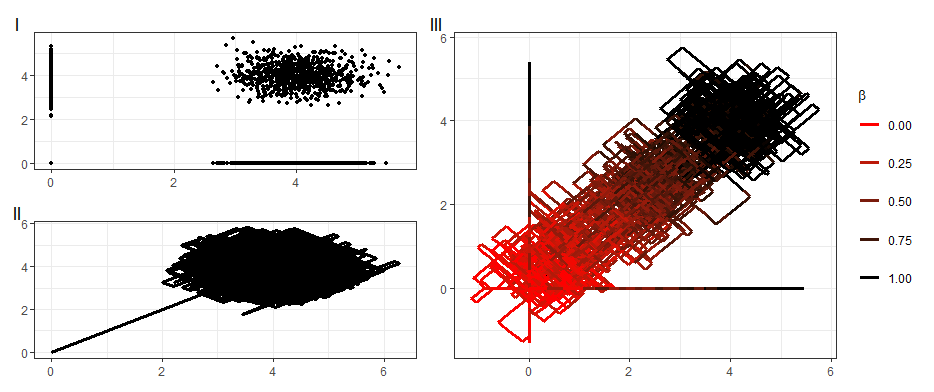}
    \caption{Sampling from \eqref{eq:spikeslab} with $m=4$. I. Exact samples from the spike-slab distribution. II. Standard Zig-Zag for $10^4$ event-times. III. Tempered Zig-Zag trajectories with $\alpha = 0.5$ for $10^4$ event-times.}
    \label{fig:spikeslab}
\end{figure}

Figure \ref{fig:spikeslab} (rightmost panel) displays the dynamics of the tempered Zig-Zag for this problem for the choice of $m=4$. Note that the standard transdimensional Zig-Zag process (bottom left) becomes stuck in a single model whilst its tempered counterpart is able to transition easily between models and thus visit the required areas for which each coordinate is equal to zero.

\begin{table}[!htbp] \centering 
  \caption{Absolute error of marginal mean and probability of inclusion for $X_1$ (averaged over 10 simulations for increasing $m$ in \eqref{eq:spikeslab}).} 
  \label{tab:spikeslab} 
\begin{tabular}{@{\extracolsep{5pt}} ll|cccccc} 
\\[-1.8ex]\hline 
\hline \\[-1.8ex] 
&&&& $\text{Mean Absolute Error} $&&&\\
& & $m=0$ &  $m=1$ & $m=2$ & $m=3$ & $m=4$ \\ \hline
$\mathbb{E}[X_1]$ & Zig-Zag & $\mathbf{0.005}$ & $\mathbf{0.018}$ & $0.633$ & $1.498$ & $1.998$\\ 
& Zig-Zag CT & $0.007$ & $0.025$ & $\mathbf{0.022}$ & $\mathbf{0.047}$ & $\mathbf{0.214}$ \\ 
$\mathbb{P}(|X_1| > 0)$ & Zig-Zag & $\mathbf{0.009}$ & $\mathbf{0.012}$ & $0.322$ & $0.500$ & $0.500$\\ 
& Zig-Zag CT & $0.010$ & $0.023$ & $\mathbf{0.008}$ & $\mathbf{0.018}$ & $\mathbf{0.055}$ \\ 
\hline \\[-1.8ex] 

\\[-1.8ex] 
\end{tabular} 
\end{table} 

Table \ref{tab:spikeslab} gives the absolute error for the Monte Carlo estimates of the marginal inclusion probabilities and marginal means. For lower $m$, the standard and tempered Zig-Zag perform similarly as the mode of the slab is close enough to ensure regular crossing of the zero-axis for the PDMP trajectories. However, as $m$ increases, the marginal probabilities of inclusion tend to 1 and the standard Zig-Zag process becomes stuck in a single model.

\subsection{Boltzmann machine relaxation}
Following \cite{Nemeth2019} and \cite{Graham2017}, the final example considers sampling a continuous relaxation of a Boltzmann machine distribution. Full details surrounding the derivation of the distribution can be found in \cite[Supplementary Material, Section D]{Nemeth2019}, though for the considered example the target density on $\mathbb{R}^{d_r}$ is of the form
\[q(\bm{x}) = \frac{2^{d_b}}{(2\pi)^{d_r/2}Z_b \exp(\frac{1}{2}{\rm Tr}(\bm{D}))}\exp\left(-\frac{1}{2}\bm{x}^\top \bm{x}\right) \prod_{k=1}^{d_b}\cosh(\bm{q}_k^\top \bm{x} + b_k),\]
where $\bm{q}_k$ denotes the $k$-th row of $\bm{Q}$, which is a $d_b \times d_r$ matrix such that $\bm{Q}\bm{Q}^\top = \bm{W} + \bm{D}$, and $\bm{D}$ is an arbitrary diagonal matrix that ensures $\bm{W} + \bm{D}$ is positive semi-definite. The above is a continuous relaxation of the Boltzmann Machine distribution on $\{-1,1\}^{d_b}$ with probability mass function 
\[q(\mathbf{s}) = Z_b^{-1}\exp\left(\frac{1}{2}\bm{s}^\top \bm{W} \bm s  + \bm{s}^\top \bm{b} \right). \] 
The moments up to second order of the relaxation and the original (discrete) Boltzmann Machine distribution are related via $\mathbb{E}[\bm{X}] = \bm{Q}^\top \mathbb{E}[\bm{S}]$ and $\mathbb{E}[\bm{X}\bm{X}^\top] = \bm{Q}^\top \mathbb{E}[\bm{S}\bm{S}^\top] + \bm{I}$. 
We employ a similar experimental setup as in \cite{Nemeth2019}, namely a $28$-dimensional example which allows for exact computation via enumeration of the moments of $\bm{S}$,  and hence in turn of $\bm{X}$ (the latter being useful for evaluating sampler performance). Table \ref{tab:boltz} displays the results. We find that the best performance is given by the Zig-Zag sampler with $\alpha = 0.2$ which far outperforms the standard Zig-Zag $\alpha=1$ and importance sampling version of the sampler. For these experiments the tuning of $\kappa$ is notably sub-optimal as the proportion of time spent at zero is not fully controlled by $\alpha$. Despite this, the samplers were able to outperform the importance sampling approach where $\alpha=0$ and the standard Zig-Zag. Further details on the experiment and specification of $\kappa$ may be found in the supplementary material. 

\begin{table}[!htbp] \centering 
  \caption{Average root-mean-square error of the first and second moments of the Boltzmann machine relaxation averaged over 20 simulations reported to 3 decimal places. } 
  \label{tab:boltz} 
\begin{tabular}{@{\extracolsep{5pt}} ccc|cc|c} 
\\[-1.8ex]\hline 
\hline \\[-1.8ex]
&&& \multicolumn{2}{c|}{Average RMSE} & Thinning \\
Method & $\alpha$ & $\omega(\T=1)$ & $\mathbb{E} [X_k] $ & $\mathbb{E}[X_k^2]$ & efficiency \\
\hline \\[-1.8ex] 
Zig-Zag & $1$ & $1$ & $1.304$ & $2.456$ & $0.146$\\ 
Zig-Zag CT & $0.700$ & $0.632$ & $0.515$ & $1.718$ & $0.187$\\ 
 & $0.500$ & $0.417$ & $0.493$ & $1.563$ & $0.219$\\ 
 & $0.300$ & $0.231$ & $0.417$ & $1.890$ & $0.251$\\ 
 & $0.200$ & $0.145$ & $\bm{0.296}$ & $\bm{1.089}$ & $0.267$ \\ 
 & $0.100$ & $0.076$ & $0.594$ & $2.643$ & $0.284$\\ 
Zig-Zag CT (IS) & $0$ & $0$ & $0.566$ & $3.580$ & $\bm{0.505}$\\ 
\hline \\[-1.8ex] 
\end{tabular} 
\end{table} 

\section{Discussion}
We present a general strategy to improve the performance of PDMP samplers on challenging targets. The approach uses an extended distribution that uses an inverse temperature which interpolates between a challenging distribution of interest and a tractable base distribution. At lower values of $\T$ the mixing will improve and we allow a pointmass at inverse at $\T=1$ to attain exact samples from the distribution. As a proof of concept, a numerical study of these ideas surrounding the Zig-Zag sampler was employed. This revealed that there are considerable benefits to the proposed extensions, as well as to introducing the point mass at $\T = 1$. We compared our tempering with a point-mass extension of the Zig-Zag to a version based on continuously tempered estimates that are given using importance sampling ideas from \cite{Graham2017}. The importance sampling approach requires restrictive choice of $\kappa$ and gave inferior performance. Our approach may be readily applied to other PDMP samplers and can incorporate assets of sampling with a PDMP such as sampling transdimensional spaces and using subsampling methods. Another avenue of research is in incorporating direct simulation from $q_0$ when the sampler hits $\T=0$ --- such moves would be particularly useful when $q_0$ is a tractable multimodal distribution.  


\bibliographystyle{abbrvnat}
\bibliography{biblio}

\end{document}


\maketitle

\vspace{-1.5cm}

\section{Proof of Theorem 1}
The measure $q(\bm{x},\T)p(\T)d\bm{x} d\T$ has a density on the open set $\mathbb{R}^{d} \times [0,1)$. We will sample it using the Zig-Zag process on the extended space $E_0 = (\mathbb{R}^{d}\times [0,1)) \times \{-1,1\}^{d+1}$.
On the other hand, the measure $\delta_{\T=1} q(\bm{x}) d\bm{x}$ is essentially a density on $\mathbb{R}^d$. We will sample it using Zig-Zag on the extended space $F = \mathbb{R}^d\times \{-1,1\}^d$.

Following the construction of \cite{chevallier2021pdmp,chevallier2020reversible}, we "stitch" $E_0$ and $F$ together through an active boundary. Let $B^{in}$ be the "entrance" boundary at the temperature $\T = 1$: $B^{in} = (\mathbb{R}^{d}\times \{1\}) \times (\{-1,1\}^{d} \times \{-1\})$, and let $B^{out}$ be the "exit" boundary $B^{out} = (\mathbb{R}^{d}\times \{1\}) \times (\{-1,1\}^{d} \times \{1\})$.

The process $Z_t$ is as follows: when in $E_0$, should it hit the boundary $B^{out}$ at time $t^-$, i.e. $Z_{t^-} \in B^{out}$, it jumps to $F$ at time $t$: $Z_t \in F$. Note that the process $Z_t$ never enters $B^{out}$. When in $F$, the process jumps back to  the entrance boundary $B^{in}$ with rate $\eta(\bm{z})$. The state space is:
\[
    E = E_0 \cup B^{in} \cup F.
\]
More precisely, if $Z_{t^-} \in B^{out}$, then $Z_t = g(Z_{t^-})$ with $g$ being the projection that removes the temperature coordinate and velocity:
\begin{align*}
    g : B^{out} &\rightarrow F \\
    (\bm{x},1,\bm{v},1) &\mapsto (\bm{x},\bm{v}).
\end{align*}
Conversely, if the process jumps from $F$ to $B^{in}$ at time $t$ then $Z_t = f(Z_{t^-})$ with 
\begin{align*}
    f : F &\rightarrow B^{in} \\
    (\bm{x},\bm{v}) &\mapsto (\bm{x},1,\bm{v},-1).
\end{align*}

With this construction, we use Theorem 2 of \cite{chevallier2021pdmp} and follow the proof of Theorem 3 of \cite{chevallier2021pdmp} for our setting to show that $\omega$ will be invariant for the process if     
    \[
        \eta(\bm{x}) =\frac{q(\bm{x},1) \kappa(1^-)}{q(\bm{x})\kappa(1)} \frac{1- \alpha}{2\alpha} = \frac{\kappa(1^-)}{\kappa(1)} \frac{1- \alpha}{2\alpha}.
    \]

Intuitively, this result is obtained by choosing an $\eta$
that balances the flows $E_0 \cup B^{in} \rightarrow F$ and $F \rightarrow E_0\cup B^{in}$.
Starting from the target distribution, the amount of mass that flows through a point $\bm{z} = (\bm{x},1,\bm{v},1) \in B^{out}$ to $(\bm{x},\bm{v}) \in F$ during a time interval of length $dt$ is $\frac{1}{2^{d+1}}q(\bm{x},1)\kappa(1^-) (1-\alpha) dt$. 
Conversely, the amount of mass that flows through a point $\bm{z} = (\bm{x},\bm{v}) \in F$ to $(\bm{x},1,\bm{v},-1) \in E_0$ is $\frac{1}{2^d}q(\bm{x}) \kappa(1) \alpha \eta(\bm{x}) dt$. Hence, the previous choice of $\eta$ balances the flows out.

\section{Continuously-tempered Zig-Zag rates}
When sampling in $E_0$, the Zig-Zag process has a rate for each component of $\bm{x}$,
\begin{align}
\label{eq:zigzagRates}
\lambda_{\bm{x}^j}(\bm{z}) = \max(0, -\bm{v}^j\partial_{\bm{x}^j}\log q(\bm{x},\T)) \quad j=1,\dots, d,   
\end{align}
and an additional rate for the inverse temperature $\T$,
$$
\lambda_{\T}(\bm{z}) = \max(0, -\bm{v}^{d+1}(\partial_{\T}\log q(\bm{x},\T) + \partial_{\T}\log \kappa(\T) ).
$$
When sampling in $F$, there are $d$ rates 
\begin{align*}
\lambda_{\bm{x}^j}(\bm{z}) = \max(0, -\bm{v}^j\partial_{\bm{x}^j}\log q(\bm{x})) \quad j=1,\dots, d,   
\end{align*}
which correspond to those given in \eqref{eq:zigzagRates} since by definition $q(\bm{x},\T=1) = q(\bm{x})$. 

\section{Simulation via thinning}
The main practical challenge with simulating the Zig-Zag process, for example with Algorithm 1, is simulating the event times. Whilst the event times depend on the state, as the state-dynamics are deterministic until the next event, these can be re-expressed as rates that depend only on time. To see this consider the $i$th event, and assume that we are at time $t$ and the current state is $\bm{z}_t$. Until there is the next event (which could be any of the $d$ possible events), the state will evolve as $\bm{z}_{t+h}=(\bm{x}_t+h\bm{v}_t,\bm{v}_t)$, thus the rate until the next event of type $i$ will be
\[
\tilde{\lambda}_i(h)=\lambda_i(\bm{z}_{t+h})=\max(0,\bm{v}_t^i \partial_{\bm{x}^i} U(\bm{x}_t+h\bm{v}_t)).
\]
Simulating a Zig-Zag process thus requires simulating events of an inhomogeneous Poisson processes of rates $\tilde{\lambda}_i(t)$. We sample a random variable $u$ uniformly in $[0,1]$ and the next event time is the time $t$ such that:
\begin{equation}
    \int_0^t \tilde{\lambda}_i(s) ds = -\log(u). 
\end{equation}
In practice, solving this equation analytically is only possible for a restricted class of rate $\tilde{\lambda}$, such as rates that are piecewise constant or piecewise linear functions of time. Where we can not simulate event times directly, we can use an approach  called \textit{thinning}. 

We find an upper rate ${\lambda}^+_i(s)$ such that $\tilde{\lambda}_i(s) \leq \lambda^+_i(s)$ for all $s$, and such that we can simulate events at rate ${\lambda}^+_i$ directly. We then propose events at this larger rate, and accept each proposed event  with probability $\tilde{\lambda}_i(t) / \lambda^+_i(t)$.

The thinning method requires computing an upper bound $\bar\lambda_i$. The computational efficiency of the resulting algorithm for simulating the Zig-Zag process is directly related to the quality (tightness) of the upper bound: a loose upper bound will lead to many rejections and therefore wasted simulation effort. We note that one useful approach for constructing appropriate upper bounds in Lemma 1 below. This approach has been used many times \citep{bierkens2020boomerang, bierkens2021sticky, Bouchard-Cote2018,  chevallier2020reversible} to construct thinning bounds and is repeated here for completeness ---  further details may be found in the derivation from Section 3.3 of \cite{Bierkens2019}.

\begin{lemma}
Suppose there exists a matrix $M = (M_{ij})_{i,j = 1}^d\in \mathbb{R}^{d\times d}$ such that for every $\bm{x}\in \mathbb{R}^d$ and element of the Hessian $H(\bm{x}) = (-\partial_i \partial_j \log q(\bm{x}))_{i,j = 1}^d$, we have $|H(\bm{x})_{i,j}|\leq M_{i,j}$ then the following linear bound 
$$
\lambda_i(s) = \max(0, -\bm{v}^i\partial_{\bm{x}^i}\log q(\bm{x} + s\bm{v})) \leq \max(0, a_i + b_is)
$$
where $a_i = -\bm{v}^i\partial_{\bm{x}^i}\log q(\bm{x})$ and $b_i = \sum_{j=1}^dM_{ij}$.  
\end{lemma}
\begin{proof}
The following in-time bound holds $$
\frac{d}{ds}\left[-\bm{v}^i\partial_{\bm{x}^i}\log q(\bm{x} + s\bm{v})\right] = -\bm{v}^i\sum_{j=1}^d \bm{v}^j\partial_{\bm{x}^j}\partial_{\bm{x}^i}\log q(\bm{x} + s\bm{v}) \leq \sum_{j=1}^d M_{ij} 
$$
from which, $\max\left(0, -\bm{v}^i\partial_{\bm{x}^i}\log q(\bm{x} + s\bm{v})\right) \leq \max\left(0, -\bm{v}^i\partial_{\bm{x}^i}\log q(\bm{x}) + s\sum_{j=1}^d M_{ij} \right)$.
\end{proof}

We note also that numerical approaches to simulating the event times have been proposed \cite{cotter2020nuzz}.

\section{Thinning bounds for geometric tempering}
\label{sec:bound}
If one can use Lemma 1 to simulate the Zig-Zag at inverse temperature $\T=0$ and $\T=1$, then implementing thinning for the geometrically tempered target is trivial. This approach assumes standard geometric tempering $q(\bm{x},\T) = q_0(\bm{x})^{1-\T}q(\bm{x})^\T$ and $\kappa(\T) = \exp(-\sum_{k=0}^m \psi_k\T^k)$ where $\psi_k\in\mathbb{R}$ are constants chosen according to Section 3.4. 

Suppose we have matrices $M$ and $M^{0}$ which bound the Hessians of the target $H(\bm{x})$ and base $H_{0}(\bm{x})$ distributions. 

\textbf{The rate function for the movement in the $\bm{x}^j$ coordinate} will depend on 
\begin{align*}
    \partial_{\bm{x}^j} \log q(\bm{x},\T) &=  \T\partial_{\bm{x}^j}\log q(\bm{x}) +(1-\T)\partial_{\bm{x}^j}\log q_0(\bm{x}).
\end{align*}
The following bound $\lambda(s)\leq \bar{\lambda}(s)$ applies 
\begin{align*}
    \bar{\lambda}(s) &= \max(0, (\T+\bm{v}^{d+1}s)(a_q + b_qs) + (1-(\T +\bm{v}^{d+1}s))(a_{q_0} + b_{q_0}s))
\end{align*}
where 
\begin{align*}
    a_q &= -\bm{v}^j\partial_{\bm{x}^j}\log q(\bm{x}),  & a_{q_0} &= -\bm{v}^j\partial_{\bm{x}^j}\log q_0(\bm{x})\\
    b_q &= \sum_{j=1}^dM_{ji}  &b_{q_0} &= \sum_{j=1}^dM_{ji}^{0}.
\end{align*}
Further simplification gives,
\begin{align}
    \label{eq:xbound}
    \bar{\lambda}(s)= \max(0, a + bs + cs^2)
\end{align}
where $a = \T a_q + (1-\T)a_{q_0},~~b = \bm{v}^{d+1}(a_q - a_{q_0})+ \T b_q + (1-\T)b_{q_0},~~\text{ and } c = \bm{v}^{d+1}(b_q - b_{q_0})$. 

\textbf{The rate function for the movement in the $\T$ coordinate} will depend on 
\begin{align*}
    \partial_{\beta} \left[\log \kappa(\T) + \log q(\bm{x},\T)\right] &=  -\sum_{k=1}^{m-1}k\psi_k\T^{k-1} + \log q(\bm{x}) - \log q_0(\bm{x}).
\end{align*}
The following in-time bound holds for $q$ and an analogous bound holds for $q_0$
\begin{align*}
    \frac{d}{ds^2}\left[-\bm{v}^{d+1}\log q(\bm{x} + s\bm{v})\right] &= \frac{d}{ds}\left[-\bm{v}^{d+1}\sum_{j=1}^d \bm{v}^j\partial_{\bm{x}^j}\log q(\bm{x} + s\bm{v}) \right]\\
    &= -\bm{v}^{d+1}\sum_{i=1}^d\bm{v}^i\sum_{j=1}^d \bm{v}^j\partial_{\bm{x}^i}\partial_{\bm{x}^j}\log q(\bm{x} + s\bm{v})\\
    &\leq \sum_{i=1}^d\sum_{j=1}^d M_{ij},
\end{align*}

from which, 
$$-\bm{v}^{d+1}\log q(\bm{x} + s\bm{v}) \leq a_q + b_qs + c_qs^2$$ 
where 
\begin{align*}
    a_q = -\bm{v}^{d+1}\log q(\bm{x}),  \qquad b_q = -\bm{v}^{d+1}\sum_{j=1}^d \bm{v}^j\partial_{\bm{x}^j}\log q(\bm{x}), \qquad c_q= \frac{1}{2}\sum_{i=1}^d\sum_{j=1}^d M_{ij}.
\end{align*}
The rate function for the inverse temperature $\T$ may be bounded by
\begin{align}
\label{eq:Tbound}
    \bar{\lambda}(s)= \max(0, -\sum_{k=1}^{m-1}k\psi_k(\T+s\bm{v}^{d+1})^{k-1} + a_q + b_qs + c_qs^2 + a_{q+0} + b_{q_0}s + c_{q_0}s^2).
\end{align}

The rates from \eqref{eq:xbound} and \eqref{eq:Tbound} are polynomial in $s$, so thinning can be implemented using the approach of \cite{sutton2021concave}.

\section{Thinning in the Examples}

In Examples 1 and 3, we use geometric tempering from an approximating multivariate-Gaussian distribution. Thinning is implemented using the arguments from section \ref{sec:bound} and bounds on the Hessians for the base and target distributions. 

\subsection{Hessian bound for a multivariate Gaussian}
One common choice for a base distribution is a $d$-dimensional multivariate Gaussian. Here
$$
q(\bm{x}) = \frac{1}{\sqrt{(2\pi)^d}\Sigma}\exp\left(-\frac{1}{2}(\bm{x} - \bm{\mu})^\top\Sigma^{-1}(\bm{x} - \bm{\mu})\right)
$$
where the Hessian is $H(\bm{x}) = (-\partial_i \partial_j \log q(\bm{x}))_{i,j = 1}^d$ so we have the upper-bound $|H(\bm{x})| \leq M$ where $M_{ij} = |\Sigma^{-1}_{ij}|$.

\subsection{Hessian bound for mixture of Gaussians}

The target has un-normalised density,
$$
q(\bm{x}) = \sum_{k=1}^K\exp\left(-\frac{1}{2\sigma^2}(\bm{x} - \bm{\mu}_k)^\top(\bm{x} - \bm{\mu}_k)\right),
$$
Following the argument in Section \ref{sec:bound}, it suffices to find a matrix that bounds the Hessian of $\log q(x)$. Since $q$ is the mixture of independent Gaussians it also follows that $H(\bm{x})_{i,j} = 0$ for $i\neq j$.  

Let $\phi_k(x) = \exp\left(-\frac{1}{2\sigma^2}(\bm{x} - \bm{\mu}_k)^\top(\bm{x} - \bm{\mu}_k)\right)$ then,
\[
    \partial_{\bm{x}^i} \log q(x) = \frac{\sum_{k=1}^K\phi_k(\bm{x})\frac{1}{\sigma^2}(\bm{x}^i-\bm{\mu}_k^i)}{ \sum_{k=1}^K\phi_k(\bm{x})} = \frac{1}{\sigma^2}\left(\bm{x}^i - \frac{1}{q(\bm{x})} \sum_{k=1}^K \bm{\mu}_k^i \phi_k(\bm{x})\right)
\]
with second derivative,
\begin{align*}
    \partial_{\bm{x}^i}\partial_{\bm{x}^i} \log q(\bm{x}) &= \frac{1}{\sigma^2}\left(1 + \frac{1}{\sigma^2} \left[\sum_{k=1}^K (\bm{\mu}_k^i)^2\frac{\phi_k(\bm{x})}{q(\bm{x})} - \left(\sum_{k=1}^K \bm{\mu}_k^i \frac{\phi_k(\bm{x})}{q(\bm{x})}\right)^2\right] \right) \\
    &\leq \frac{1}{\sigma^2}\left(1 + \frac{1}{\sigma^2} \frac{1}{4}(M^i - m^i)^2 \right)
\end{align*}
where $M^i = \max_k \{\bm{\mu}^i_k\}$ and $m^i = \min_k \{\bm{\mu}_k^i\}$ and the bound follows from Popovicius inequality. We have the bound $|H(\bm{x})|\leq M$ where $M_{ij} = 0$ for $i\neq j$ and $M_{i,i} = \frac{1}{\sigma^2}\left(1 + \frac{1}{\sigma^2} \frac{1}{4}(M^i - m^i)^2 \right)$ otherwise. 


\subsection{Boltzmann machine relaxation}

The target has un-normalised density,

\[q(\bm{x}) = \frac{2^{d_b}}{(2\pi)^{d_r/2}Z_b \exp(\frac{1}{2}{\rm Tr}(\bm{D}))}\exp\left(-\frac{1}{2}\bm{x}^\top \bm{x}\right) \prod_{k=1}^{d_b}\cosh(\bm{q}_k^\top \bm{x} + b_k).\]

Following the argument from Section 2, it suffices to find a bound on the Hessian matrix. The first order derivatives are
$$
-\nabla_{\bm{x}} \log q(\bm{x}) = \bm{x} - \sum_{k=1}^{d_b} \bm{q}_k\tanh(\bm{q}_k^\top \bm{x} + b_k)
$$
and the Hessian matrix is
$$
H(\bm{x}) = \bm{I} - \sum_{k=1}^{d_b} \bm{q}_k\bm{q}_k^{\top}\text{sech}^2(\bm{q}_k^\top \bm{x} + b_k).
$$
As $0\leq \text{sech}^2(t) \leq 1$ we have 
$$
H(\bm{x}) \leq \bm{I} - \sum_{k=1}^{d_b} \min[0, \bm{q}_k\bm{q}_k^{\top}] = M^+, \qquad H(\bm{x}) \geq  -\sum_{k=1}^{d_b} \max[0, \bm{q}_k\bm{q}_k^{\top}] = M^-.
$$
We have the bound $|H(\bm{x})|\leq M$ where $M = \max(|M^+|, |M^-|)$. Thinning may be implemented using the argument from Section \ref{sec:bound}.

\subsection{Transdimensional example}
The transdimensional example does not use geometric tempering, but the bounds are simple to construct. The tempering is 
$$
q(\bm{x}, \T) = \prod_{i=1}^2(w\phi(\bm{x}^i; m\T, \sigma^2) + (1-w)\delta_0(\bm{x}^i)),
$$
where $\phi$ is the normal density function. For this example, we took $\kappa(\T) = 1$. 

Variables that are not stuck at the pointmass are simulated according to the slab $\mathcal{N}(\mu\T, \sigma^2)$ distribution. With gradient $\partial_{\bm{x}^i}\log q(\bm{x},\T) = \frac{1}{\sigma^2}(\bm{x}^i -m\T)$, their event rate is
$$
\lambda_{\bm{x}^i}(s) = \max\left(0, \frac{\bm{v}^i}{\sigma^2}((\bm{x}^i+s\bm{v}^i) - \mu(\T+s\bm{v}^{d+1})\right)
$$
which is a linear function in $s$ that may be simulated exactly using the methods of \cite{sutton2021concave}. Following \cite{chevallier2020reversible} the rate to reintroduce a variable $\bm{x}^i=0$ is given by
$$
\frac{w}{(1-w)}\phi(0; m\T), \sigma^2) = \frac{w}{(1-w)}\frac{1}{\sqrt{2\pi\sigma^2}}\exp\left(-\frac{m^2}{2\sigma^2}\T^2\right) 
$$
which admits thinning using $\bar{\lambda}(s) = \frac{w}{(1-w)}\frac{1}{\sqrt{2\pi\sigma^2}}$. The rate for the temperature (when not stuck at $\T=1$) is
$$
\lambda(s) = \max\left(0, \frac{m\bm{v}^{d+1}}{\sigma^2}\sum_{i: |\bm{x}^i|>0}\left(\bm{x}^i+s\bm{v}^i - m(\T+s\bm{v}^{d+1})\right)\right),
$$
which is a linear function of $s$ that may be simulated exactly using the methods of \cite{sutton2021concave}. 

\section{Additional simulation details}

\subsection{Mixture of Gaussians}

For the Mixture of Gaussians the location of the means are given below (stated to 2 decimal places).  

\begin{table}[!htbp] \centering 
  \caption{Location of the means for the Gaussian mixture } 
  \label{} 
\begin{tabular}{@{\extracolsep{5pt}} lccccc} 
\\[-1.8ex]\hline 
\hline \\[-1.8ex] 
$\bm{\mu}^1$ & $2.66$ & $5.73$ & $2.02$ & $9.45$ & $6.29$ \\ 
$\bm{\mu}^2$ & $3.72$ & $9.08$ & $8.98$ & $6.61$ & $0.62$ \\ 
\hline \\[-1.8ex] 
\end{tabular} 
\end{table} 

For $\sigma^2 = 0.5$ the distance between most local modes is greater than 15 standard deviations with the minimum distance being 14.85 standard deviations from it's closest neighbour. This presents a challenging posterior for sampling as seen in Figure 1 of the main paper. For this example the exact first and second moment of the Gaussian mixture can be calculated exactly and is stated in Table \ref{tab:exactMOM} below. 

\begin{table}[!htbp] \centering 
  \caption{Table of exact first and second moments of the Gaussian mixture model} 
  \label{tab:exactMOM} 
\begin{tabular}{@{\extracolsep{5pt}} lcc} 
\\[-1.8ex]\hline 
\hline \\[-1.8ex] 
 & $X_1$ & $X_2$ \\ 
 \hline\\[-1.2ex]
$\mathbb{E}[X_k]$ & $5.228$ & $5.803$ \\ 
$\mathbb{E}[X_k^2]$ & $34.751$ & $44.418$ \\ 
\hline \\[-1.2ex] 
\end{tabular} 
\end{table} 

Boxplots showing the variability of estimated first and second moments of $X_1$ show the performance improvement given using tempering $\alpha\neq 1$ and using a point-mass $\alpha\neq 0$. 

\begin{figure}[!htbp]
    \centering
    \includegraphics[width=.8\textwidth]{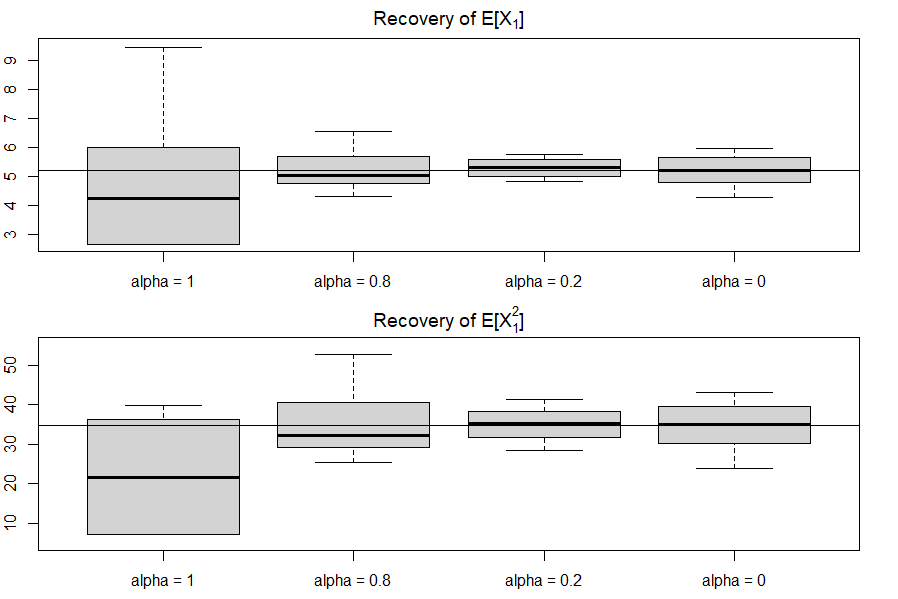}
    \caption{Recovery of the first and second moments of $X_1$ for the Gaussian mixture model}
    \label{fig:bprecovery}
\end{figure}

\subsection{Tuning of $\kappa(\T)$ in experiments}
For the tempered Zig-Zag, $\kappa(\T)$ is chosen to approximate $Z(\T)$. This quantity may be estimated using many approaches outlined in \cite{Gelman1998}. For our experiments, we use numerical integration as described in Section 2.3 of \cite{Gelman1998} further details may be found therein. We use the trapezoidal rule to estimate
$$
\widehat{\log Z(\T_{(k)})} = \frac{1}{2}\sum_{j=1}^{k-1}(\T_{(j+1)}-\T_{(j)})(\bar{U}_{(j)} + \bar{U}_{(j+1)}),
$$
where $\bar{U}_{(j)} = \mathbb{E}\left[ \partial_{\T} \log q(\bm{x}, \T) \mid \T = \T_{(j)}\right]$ is estimated using Monte Carlo and the values of $\T$ are ordered so that $\T_{(1)}<\T_{(2)}<\dots < \T_{(k)} \leq \dots<\T_{(n)}$. In practice, either a finite grid of fixed values $\T$ from 0 to 1 can be used to construct this estimate or a prior run of the Zig-Zag with an uniformed choice $\kappa(\T)\propto 1$ may be used to obtain these samples. We may then fit the regression model
$$
\widehat{\log Z}(\T) \approx \log \kappa(\T),
$$
to specify the polynomial terms in $\kappa(\T)$.

In Example 1 (Gaussian mixture model), all methods were run for 50,000 events and the first 40\% was used as burnin and tuning of $\kappa$. In the initial, burnin sampling we specified $\kappa(\T)\propto 1$ with $\alpha = 0$. The tempered Zig-Zag samplers then used the events from this burnin process to construct an estimate of $\widehat{\log Z}(\T)$. The samplers were then run for the remaining 30,000 event times and the estimated first and second moments were recorded.

In Example 2 (the transdimensional example), we fix $\kappa(\T)=1$ because the marginal distribution for the inverse temperature $\T$ was sufficiently close to uniformly distributed.

In Example 3 (Boltzmann machine relaxation), a finite grid of 15 $\T$ values equally spaced from 0.01 to 0.99 were used to form the construction of $\widehat{\log Z}(\T)$. The associated choice of $\kappa$ was used for all continuously tempered methods. For $\alpha=0$ tempering with importance sampling, we used $\kappa(\T) = \log \xi^{1-\T}$ where $\xi$ was specified using a variational Gaussian approximation to the target as in \cite{Graham2017} and \cite{Nemeth2019}. 

\subsection{Computational resources}

All experiments were implemented using the code accompanying the supplementary material. The multiple runs required for the simulation study were implemented in parallel using high performance computational resources. This amounted to submitting job requests for each individual replicate of the simulation studies. In each replicate the methods were given the same amount of computational resources i.e. simulated event-times. The results of the parallel runs were collected and processed to evaluate the performance of the methods --- i.e. calculation of the average root mean square error. 

\bibliographystyle{abbrvnat}
\bibliography{biblio}